\documentclass[runningheads]{llncs}
\usepackage{amsmath}
\usepackage{algorithm}
\usepackage{algpseudocode}
\usepackage{multicol}
\usepackage{multirow}
\usepackage{subcaption}
\usepackage{tabularx}
\usepackage{hhline}
\usepackage{setspace}
\usepackage{amsmath}
\usepackage{graphicx}
\graphicspath{{../rsc/}}
\usepackage{array}
\usepackage{etoolbox}
\usepackage{dblfloatfix}
\usepackage{hyperref}
\usepackage[figuresright]{rotating}
\hypersetup{
    colorlinks=true,
    filecolor=magenta,      
    urlcolor=blue,
}
\begin{document}
\title{Security Requirements of Commercial Drones for Public Authorities by Vulnerability Analysis of Applications}
\titlerunning{Security Requirements of Drones by Vulnerability Analysis of Apps}

\author{Daegeon Kim\orcidID{0000-0002-6142-9682} \and
Huy Kang Kim\orcidID{0000-0002-0760-8807}}
\authorrunning{D. Kim and H. K. Kim}

\institute{Graduate School of Information Security, Korea University, Republic of Korea\\
\email{\{dgkim0803, cenda\}@korea.ac.kr}}
\maketitle              

\begin{abstract}
Due to the ability to overcome the geospatial limitations and to the possibility to converge the various information communication technologies, the application domains and the market size of drones are increasing internationally. Public authorities in South Korean are investing for the domestic drone industry and the technological advancement as a power of innovation and growth of the country. They are also increasing the utilization of drones for various purposes. 

The South Korean government ensures the security of IT equipment introduced to the public authorities by enforcing policies such as security compatibility verification and CCTV security certification. Considering the increase of the needs of drones and the possible security effects to the organization operating them, the government needs to develop the security requirements during introducing drones, but there are no such requirements yet. 

In this paper, we inspect the vulnerabilities of drones by analyzing the applications of commercial drones made by 4 manufacturers. We also propose the minimum security requirements to resolve the vulnerabilities. We expect our work contributes to the security improvements of drones operated in public authorities.

\end{abstract}

\section{Introduction}
A drone is a sort of small flying object without an onboard human pilot. It is being adopted and operated to the broad area thanks to its characteristics such as the efficiency to the human resource requirement and the ability to overcome geographical limitations. The possibility to convergence with information communication technologies, including data collection, transmission, processing, and analysis allows continuous expansion of the applicability of drones.

A drone is mainly used to record video for reconnaissance, surveillance, management, and measurement by installing image collection equipment, but it is also used as the flight vehicle for such as transportation and disaster prevention. It is widely used for sports and leisure like racing, too. In 2018 PyeongChang winter Olympic opening ceremony, it is adopted as the mean of the arts to project three-dimensional objects in the air by applying the cluster flight control and the wireless communication technologies to drones. In the military domain, it is possible to be used as the weapon equipped guns or explosives on it and \cite{DKim2015} showed the effectiveness of the operation droned in the battlefield for surveillance.

As drone-related technologies are developed so that they can be used in various fields, the demands of drones and the size of markets are also expanding. According to \cite{JPark2018}, the drone market can be divided into commercial and military applications. The global drone market combining the two sectors will grow from \$86.8 billion in 2015 to \$143.9 billion in 2020, with an average annual growth rate of 10.64 percent. Of the total, the global market for commercial drones will reach about \$63.1 billion in 2020 from \$19.2 billion in 2016, with an average annual growth rate of 34.6 percent.

As the growth of the size of commercial drone markets accelerates, it is expected that public authorities will also expand the introduction of commercial drones that have functions and capabilities that are suitable for operating purposes. The drones used in those areas require a higher security level as they have to protect the public property and to maintain national security.

Advances in IT technologies increase the adoption of various equipment incorporating the latest technologies into public authorities at the same time require the security of the equipment. When equipment containing security functions is introduced to public authorities in South Korea, the security compliance verification system is applied. Also, for the equipment which can result in extensive security impact like CCTV, the public authorities are forced to introduce security verified products by the certificating authority applying the independent security requirement. However, even though commercial drones are widely used for security purposes so that the security impacts are high, there are no security requirements except for the cryptography area that can be applied to drones introduced in the public authorities.

In this paper, we propose the minimum security requirements that can be applied when introducing commercial drones to public authorities. For this, we analyzed vulnerabilities of the apps of four manufacturers' commercial drones and derived the requirements to prevent identified vulnerabilities.

This paper consists as follows. In section 2, we examine the researches related to the vulnerability and the security requirements of drones. The vulnerabilities of the commercial drones are analyzed in section 3, and security requirements are proposed in section 4 after categorizing the types of vulnerabilities. Section 5 concludes this paper.

\section{Related Works}
\label{section:RelatedWorks}
\subsection{Vulnerabilities of drones}
A drone has consisted of a flight object and a controller. Considering the way drones are operated such that the two components exchange data in remote communication channels, expected vulnerabilities of them can be derived. \cite{IoTFS2015} and \cite{TTAK2017} categorized representative security vulnerabilities of drones as Table~\ref{tab:DroneVulnerabilityTypes}.

\begin{table}[h]
    \centering
    \caption{Types of vulnerability of drone}
    \label{tab:DroneVulnerabilityTypes}
    \begin{tabular}{l|l}
        \hline
        \# & vulnerability types \\ \hline
        1 & cipher key hacking after hijacking drone \\
        2 & neutralization of drone (e.g., GPS jamming) \\
        3 & information leaking \\
        4 & malware infection \\
        5 & loading unauthorized device \\ \hline 
    \end{tabular}
\end{table}

The drone may contain several vulnerabilities which cause each vulnerability types. \cite{IJun2017} and \cite{DHong2017} found the vulnerabilities of AR Drone manufactured by Parrot such as deauthorizations, remote command execution, and file tampering abusing telnet and FTP service.

\cite{JValente2017} discovered the vulnerabilities of U818A manufactured by Discovery whose drones are widely re-branded as other manufacturers. The vulnerabilities include weak authentication resulting account stealing, telnet account information extortion from the password file which is set inappropriate access control, and FTP administrator privilege takeover by tampering the password file. What's interesting about the study is that using the Discovery drone's controller app, they were able to control at least 17 drones from other manufacturers controlled by different apps. They identified three commonalities from those apps, including fixed IP addresses set for the drone. From those findings, it can be assumed that many manufacturers reuse drone control apps, and the same vulnerabilities are contained in them.

There are studies of the risk assessment methodology specialized for drones. \cite{KHartmann2013} analyzed the drone hacking cases and indexed the risks in terms of confidentiality, integrity, and availability of drone's communication channel, data storage, sensor and error control component depending on the mission and the operating environment to be performed by the drone. They assessed the risk by combining the indices.

The previous studies related to the vulnerability of drones mainly focused on the categorization of security vulnerabilities and the vulnerabilities held by specific models, but it is hard to find studies that classify the vulnerabilities of the drone by analyzing vulnerabilities of multiple drone models.

\subsection{Security requirements of drones}
We can find the literature of security requirements of drones from the study of security requirements standards and the Protection Profiles (PP) for the Common Criteria (CC).

\cite{IoTFS2015} and \cite{TTAK2017} are the organization standards of the security requirement of the environment composed by a service provider utilizing drones and the consumers of the service. The standards define 6 components of the drone-based service system, including service requester, service providing organization, ground controller, and drones. To prevent the security vulnerabilities of drones in Table~\ref{tab:DroneVulnerabilityTypes}, the security requirements of each component and those of the interfaces between them are proposed classified by compulsory and optional items.

\cite{WJeon2016} defined PP of drones for the CC certification. To develop evaluation criteria which is the basis of PP, 1) 7 security threats and 3 assumptions are derived from the drone attack scenario, 2) the security objectives are set to prevent the threats, 3) 26 security functional requirements in 7 categories were developed to fulfill each objective.

Although there are security criteria established based on assumptions, theories, and threat scenarios as the mentioned literature, it is rare to find studies deriving security requirements based on the technical analysis of security vulnerabilities of drones. In this respect, the study is highly empirical.

\section{Vulnerability Analysis of Drones}
\label{section:VulnerabilityAnalysis}
\subsection{Targets}
In this study, we analyzed security vulnerabilities in control apps for Android operating systems of commercial drones from four manufacturers on the market. Detailed information of the analysis targets is shown in Table~\ref{tab:TargetInfo}. As in the table, a single dedicated drone control app allows controlling several drone models sold by a manufacturer. Thus, the security vulnerabilities of the drones analyzed through the control apps in this paper may not only exist in a particular drone model but may exist simultaneously in a variety of models.

\begin{table*}[!th]
    \centering
    \caption{Detailed information of the analysis targets}
    \label{tab:TargetInfo}
    \begin{tabular}{c|c|c|c|c}
       \hline
       manufacture & country & drone models &  control app & app version \\ \hline
       Syma & China & X5 series, X8 series & SYMA FPV & 5.2 \\ 
       Parrot & France & AR. Drone, BEBOP series & AR FreeFlight & 2.4.15 \\ 
       Yuneec & China & Typhoon series & CGO3 & 0.1 \\ 
       UDI R/C & China & U29, U31, U47 & UDIRC fpv & 2.3 \\ \hline
    \end{tabular}
\end{table*}

\subsection{Methods}
Static analysis and dynamic analysis of the apps were conducted to identify the security vulnerabilities of the control apps.

The primary function of the control app is to provide a controller with functions for drone control and data transmission. We set the objective of the static analysis to identify vulnerabilities in data transferred between the controller and the drone, in implemented services, and in the app itself.

If the behavior between the controller and the drone is analyzed in the dynamic analysis, different results can be produced depending on the model types. Also, the purpose of this paper is to analyze the security vulnerability identifiable in the control app itself, not limited to the specific drone model. Thus, we set the objective of the behavior analysis as identifying vulnerabilities in activities between third-party web servers and the controller.

\subsection{Results}
\subsubsection{Syma (SYMA FPV v5.2)}
\begin{figure*}[hbt!]
    \centering
    \includegraphics[width=0.6\textwidth]{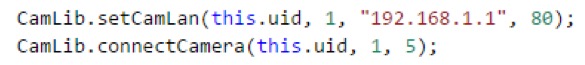}
    \caption{The static IP address of the camera in CameraManage.java}
    \label{fig:SYMA_FixedCameraIP}
\end{figure*}

A fixed IP address is assigned to the camera attached to the drone as Figure~\ref{fig:SYMA_FixedCameraIP} and no authentication is required to access and to control it. This allows all devices belonging to the same network with the camera can send and receive data to/from it.

\begin{figure*}[hbt!]
    \centering
    \includegraphics[width=1.0\textwidth]{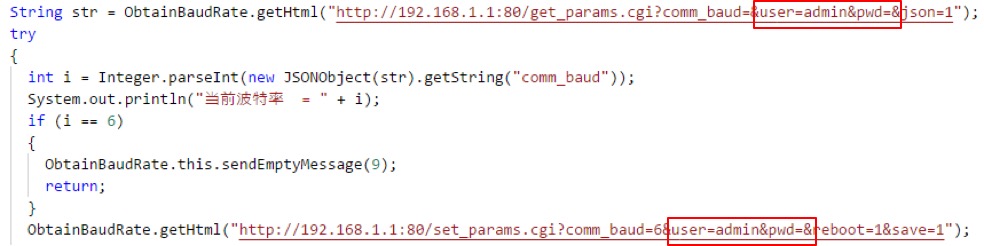}
    \caption{The plain text user ID and password included in the query string to access CGI file in ObtainBauRate.java}
    \label{fig:SYMA_UserID}
\end{figure*}

To make the interaction between the web service in the drone and the controller, Common Gateway Interface (CGI) protocol which is the standard protocol to make the interaction between application programs and web servers is used to transfer data. \cite{DYoo2007} found the vulnerability of the gateway in the home network system, which is caused by CGI protocol. In the same way, the drone manufactured by Syma also has the file named \textit{get\_params.cgi} which stores the account information and the configuration settings of the camera as well as the several web service of the drone where the configuration setting can be modified by the file named \textit{set\_params.cgi}. The vulnerability arises due to the lack of access control toward those files. Moreover, the plain text ID and password to be used to access the drone is included in the query strings set to execute CGI files as shown in Figure~\ref{fig:SYMA_UserID}.

\subsubsection{Parrot (AR FreeFlight v2.4.15)}
\begin{figure*}[bht!]
    \centering
    \includegraphics[width=1.0\textwidth]{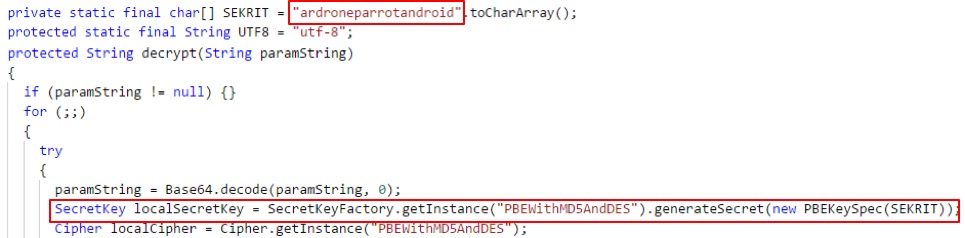}
    \caption{The initial value to generate a password based encryption key in AcademySharedPreferences.java}
    \label{fig:PARROT_EncryptionKey}
\end{figure*}

The crucial data stored in the mobile controller is encrypted by the key generated by the initial value. However, since the initial value is exposed in the app which can be extracted by decompiling the app as Figure~\ref{fig:PARROT_EncryptionKey}, one can decrypt the stored data.

\begin{figure*}[hbt!]
    \begin{tabular}{p{0.48\textwidth}p{0.48\textwidth}}
    
        \begin{subfigure}{.5\textwidth}
            \includegraphics[width=0.9\textwidth]{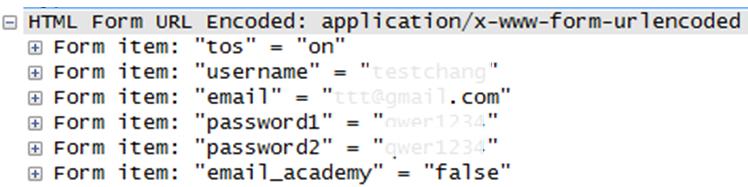}
            \caption{parrot account of an user}
        \end{subfigure}
        
        &   \multirow{2}{\hsize}[10mm]{
                \begin{subfigure}{.5\textwidth}
                    \includegraphics[width=0.9\textwidth]{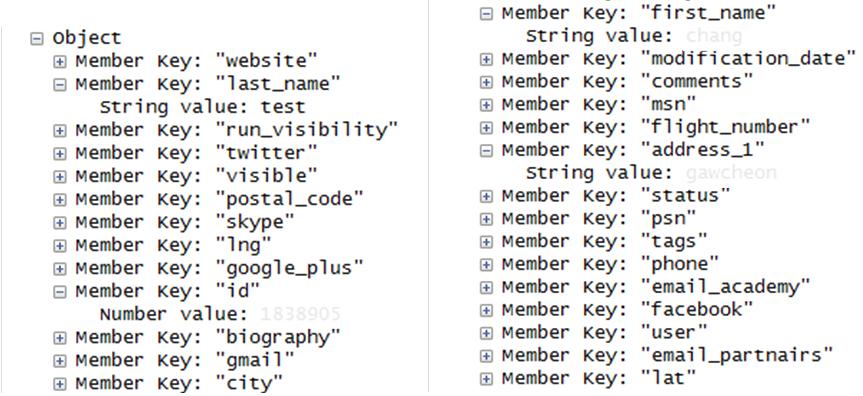}
                    \caption{other accounts and user information}
                \end{subfigure}
             }
        \\
        \begin{subfigure}{.5\textwidth}
            \includegraphics[width=0.9\textwidth]{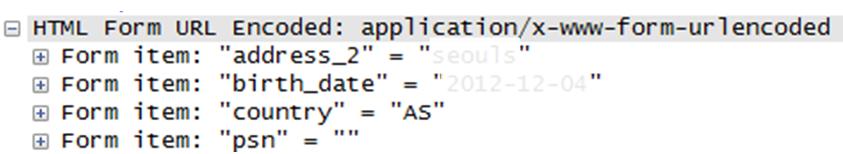}
            \caption{user information}
        \end{subfigure}
         &    
         
    \end{tabular}
    \caption{Account and user information transmitted in plain text}
    \label{fig:PARROT_PlaintextInfo}
\end{figure*}

When logging in from the mobile app to the server of the manufacture and modifying the personal information, various information is transferred in plain text as the packet captured in Figure~\ref{fig:PARROT_PlaintextInfo}.

\begin{figure*}[hbt!]
    \begin{tabular}{p{\textwidth}}
        \begin{subfigure}{\textwidth}
          \centering
          \includegraphics[width=1.0\textwidth]{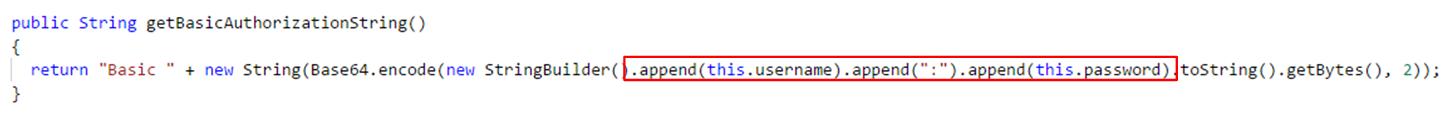}  
          \caption{AcademyCredentials.java}
        \end{subfigure}
        \\
        \\
        \begin{subfigure}{\textwidth}
          \centering
          \includegraphics[width=1.0\textwidth]{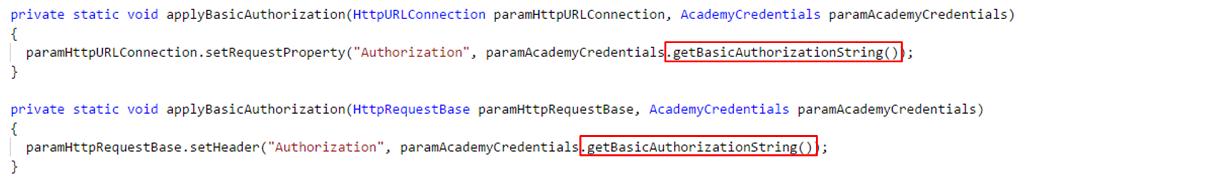}  
          \caption{Academy.java}
        \end{subfigure}
    \end{tabular}

    \caption{(a) The authentication string composed by the simple combination of username and password, (b) generating a user authorization credential using the string to access a manufacture's web server.}
    \label{fig:PARROT_AuthString}
\end{figure*}

For the user logged in to the manufacture's server, an authentication string is generated to check the right to execute the functions such as saving, modifying, and deleting personal information. However, this string is generated by merely composing the plain text username and password as in Figure~\ref{fig:PARROT_AuthString} which can be easily imitated.

\begin{figure*}[hbt!]
    \begin{tabular}{p{1.0\textwidth}}
        \begin{subfigure}{1.0\textwidth}
          \includegraphics[width=\textwidth]{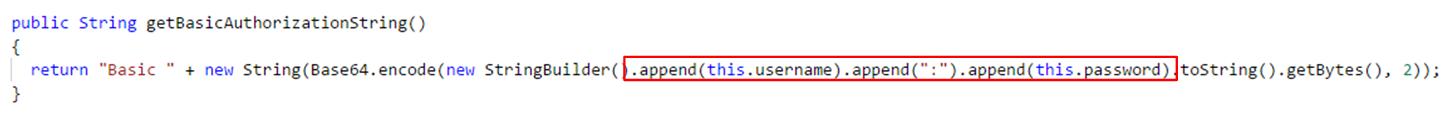}  
          \caption{FTPUtils.java}
        \end{subfigure}
        \\
        \\
        \begin{subfigure}{1.0\textwidth}
          \includegraphics[width=\textwidth]{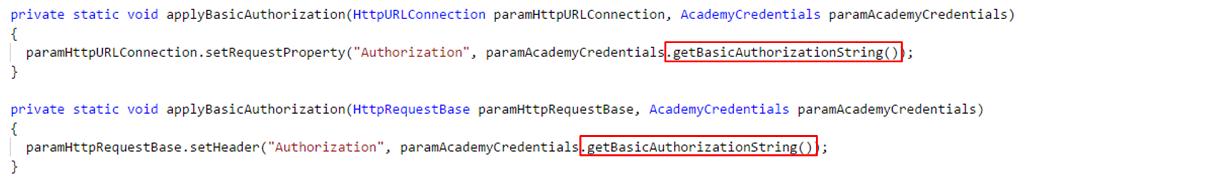}  
          \caption{FTPClient.java}
        \end{subfigure}
    \end{tabular}

    \caption{(a) Requesting a FTP connection, (b) generating the requested FTP connection as an anonymous user}
    \label{fig:PARROT_FTP}
\end{figure*}

Anonymous FTP and telnet services, known for their various vulnerabilities, have been run on the drone to upload files and to execute remote commands as shown in Figure~\ref{fig:PARROT_FTP}. This can be exploited by an attacker to upload malware to the drone and control it.

\subsubsection{Yuneec (CGO3 v0.1)}
\begin{figure*}[hbt!]
    \begin{tabular}{p{\textwidth}}
        \begin{subfigure}{\textwidth}
          \centering
          \includegraphics[width=0.6\textwidth]{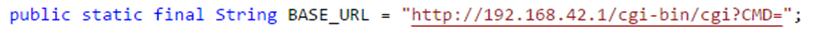}  
          \caption{CGO3BaseRequest.java}
        \end{subfigure}
        \\
        \\
        \begin{subfigure}{\textwidth}
          \centering
          \includegraphics[width=0.7\textwidth]{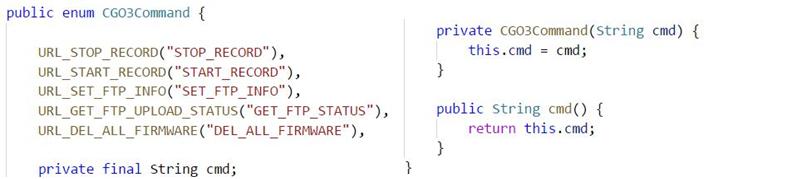}  
          \caption{CGO3Command.java}
        \end{subfigure}
        \\
        \\
        \begin{subfigure}{1.0\textwidth}
          \centering
          \includegraphics[width=1.0\textwidth]{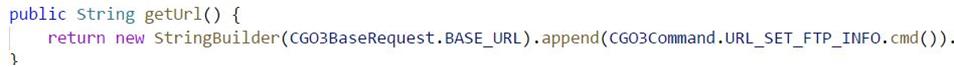}  
          \caption{StartSendRequest.java}
        \end{subfigure}
        \\
        \\
        \begin{subfigure}{1.0\textwidth}
          \centering
          \includegraphics[width=1.0\textwidth]{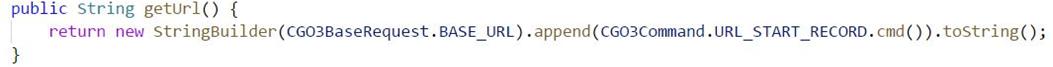}  
          \caption{StartRecordRequest.java}
        \end{subfigure}
    \end{tabular}

    \caption{(a) Base URL for CGI command, (b) sample CGI commands, and (c), (d) examples composing full URL to execute CGI command}
    \label{fig:YUNEEC1}
\end{figure*}

Similar to the case of Syma, the drone manufactured by Yuneec also assigned a fixed IP, and various CGI command files are remotely executable to control the camera and to exchange data between the controller and the drone as captured in Figure~\ref{fig:YUNEEC1}.

\begin{figure*}[hbt!]
    \centering
    \includegraphics[width=\textwidth]{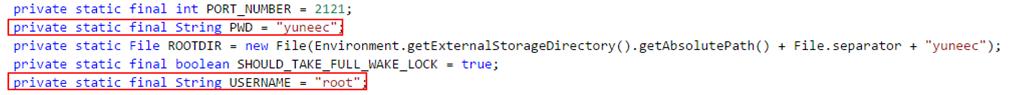}
    \caption{The hard coded FTP server access ID and password in plain text in FileSettings.java}
    \label{fig:YUNEEC2}
\end{figure*}

FTP service is running on the controller to exchange files with the drone where the username and the password to access the server are hardcoded in plain text as Figure~\ref{fig:YUNEEC2}.

\begin{figure*}[hbt!]
    \begin{tabular}{p{\textwidth}}
        \begin{subfigure}{\textwidth}
          \centering
          \includegraphics[width=0.8\textwidth]{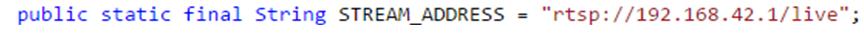}  
          \caption{CGO3.java}
        \end{subfigure}
        \\
        \\
        \begin{subfigure}{\textwidth}
          \centering
          \includegraphics[width=0.8\textwidth]{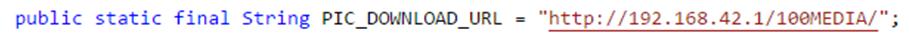}  
          \caption{HttpRequest.java}
        \end{subfigure}
    \end{tabular}

    \caption{The address and the paths to access (a) video streams of a camera, and (b) images stored on the drone}
    \label{fig:YUNEEC3}
\end{figure*}

The path to get real-time video stream recorded by the drone-attached camera and to access stored video file in it is hardcoded, and no access control method is implemented to access the path as Figure~\ref{fig:YUNEEC3} which allows the malicious capture of the video.

\begin{figure*}[hbt!]
    \centering
    \includegraphics[width=0.8\textwidth]{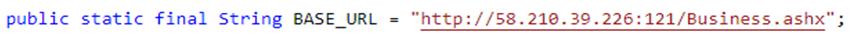}
    \caption{The vague public IP address hard coded in HttpRequest.java}
    \label{fig:YUNEEC4}
\end{figure*}

Also, the IP address not accessed by the app as well as by the drone so that the objective is unclear is included in the app as shown in Figure~\ref{fig:YUNEEC4}. This address can be abused to lead unintended access to malicious external servers.

\subsubsection{UDI R/C (UDIRC fpv v2.3)}
It was hard to conduct static analyze the source code extracted by decompiling the app of UDI R/C's drone because of obfuscation technology applied to the app. However, we were able to find the vulnerability characteristics of the drones manufactured by Discovery mentioned in \cite{JValente2017} by analyzing the behavior of the app.

\cite{JValente2017} discovered 17 different apps which can control Discovery's drones where the apps share the characteristics as below.

\begin{enumerate}
    \item The IP address of the drone is fixed to 192.168.0.1
    \item The unique string, \textit{66808000808080800c0c99}, is keep sent to 50000 port of the drone using UDP
    \item The commands to control the drone are sent to 7060, 8060, and 9060 ports of the drone using TCP.
\end{enumerate}

\begin{figure*}[hbt!]
    \centering
    \includegraphics[width=\textwidth]{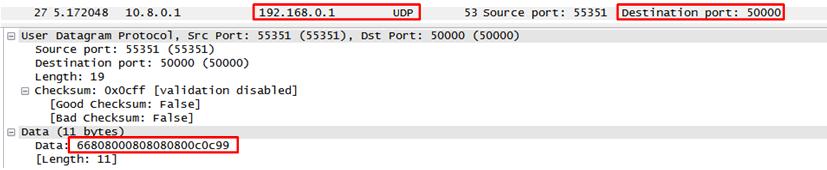}
    \caption{The distinct string sent by UDP to 50000 port of 192.168.0.1}
    \label{fig:UDI1}
\end{figure*}

\begin{figure*}[hbt!]
    \centering
    \includegraphics[width=\textwidth]{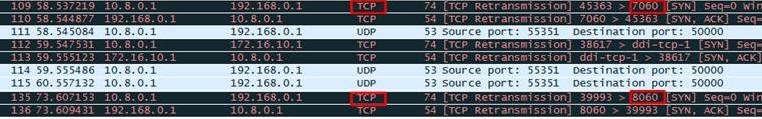}
    \caption{The TCP connections to 7060 and 8060 ports of 192.168.0.1}
    \label{fig:UDI2}
\end{figure*}

Figure~\ref{fig:UDI1} is capturing the characteristics the first two characteristics UDI R/C drone is including and Figure~\ref{fig:UDI2} shows the last characteristic it has.

\cite{JValente2017} also discovered the vulnerability to intercept the video stream and the stored information to the drone through 7060 and 8060 ports of it. It can reasonably assume that UDI R/C drones also have the same vulnerability.

\section{Proposing Security Requirements}
\subsection{Categorizing Types of Vulnerability}

\begin{sidewaystable}
    \centering
    \caption{The summery of analyzed vulnerabilities of commercial drones}
    \label{tab:VulnerabilitySummary}
    \begin{tabular}{c|l}
       \hline
       code & vulnerability \\ \hline
       Syma-1 & the static IP address assigned to the camera on the drone without access control  \\ 
       Syma-2 & open access to CGI files which are required to control a drone using CGI protocol  \\ 
       Syma-3 & exposed ID and password to access a drone in query strings of URL to access CGI command  \\ 
       Parrot-1 & exposed the plain text initial string to generate cipher key in the app  \\ 
       Parrot-2 & transmitting plain text account and user information while accessing the Parrot server  \\ 
       Parrot-3 & generating a weak authentication credential by the simple composition of plain text ID and password  \\ 
       Parrot-4 & providing vulnerable services, anonymous FTP and telnet, on the drone  \\ 
       Yuneec-1 & static IP assigned to the camera on the drone without access control  \\ 
       Yuneec-2 & open access to CGI files which are required to control the drone using CGI protocol  \\ 
       Yuneec-3 & exposed ID and password to access FTP server in the app \\ 
       Yuneec-4 & exposed file system directory paths for streaming video and storing image without access control \\ 
       Yuneec-5 & the hardcoded vague public IP address in the app  \\ 
       UDI-1 & allowing to control the drone through open ports \\ \hline
    \end{tabular}
    
    \bigskip\bigskip\bigskip  

    \centering
    \caption{The summery of analyzed vulnerabilities of commercial drones}
    \label{tab:VulnerabilityCategory}
    \begin{tabular}{c|l|l}
       \hline
       code & type & corresponding vulnerabilities\\ \hline
       V-1 & unauthorized access to file systems or devices & Syma-1$\cdotp$2, Yuneec-1$\cdotp$2$\cdotp$4, UDI-1 \\ 
       V-2 & critical information exposure in the app & Parrot-1 \\ 
       V-3 & critical data transmitted as plain text & Syma-3, Parrot-2$\cdotp$3, Yuneec-3 \\ 
       V-4 & vulnerable application services provided by a drone or a control device & Parrot-4, UDI-1  \\ 
       V-5 & unintended access to external server & Yuneec-5 \\ \hline
    \end{tabular}
    
    \bigskip\bigskip\bigskip  
        
    \caption{The security requirements of the commercial drones for the public authorities}
    \label{tab:SecurityRequirements}
    \begin{tabular}{c|l|l}
       \hline
       category & security requirement & target \\ \hline
       \multirow{2}{*}{authorization} & \#1. Authorize legitimate users during controlling the drone and using core functionality & V-1 \\ 
       & \#2. Restrict repeated access trials which are fail to be authorized & V-1 \\ \hline
       \multirow{3}{*}{\begin{tabular}[c]{@{}c@{}}critical\\ information\\ protection\end{tabular}} & \#3. Protect critical information in the app & V-2 \\ 
       & \#4. Encrypt critical data before transmitted through network & V-3 \\ 
       & \#5. Use secure encryption algorithm for data encryption & V-2$\cdotp$3 \\ \hline
       \multirow{2}{*}{\begin{tabular}[c]{@{}c@{}}access \\ control\end{tabular}} & \#6. Restrict unintended access to external servers & V-5 \\ \
       & \#7. Restrict multiple user accesses to control the drone & V-1$\cdotp$2$\cdotp$4 \\ \hline
       \multirow{2}{*}{authorization} & \#8. Restrict unnecessary services and ports  & V-4 \\ 
       & \#9. Restrict arbitrary control commands unintended by users & V-4 \\ \hline
    \end{tabular}
\end{sidewaystable}

The vulnerabilities explained in section~\ref{section:VulnerabilityAnalysis} is summarized in Table~\ref{tab:VulnerabilitySummary}. Also the types of vulnerability is categorized in Table~\ref{tab:VulnerabilityCategory} with corresponding vulnerabilities in Table~\ref{tab:VulnerabilitySummary}

\subsection{Security Requirements of The Commercial Drone}
It may be possible to require to have high-security features, including electronic warfare response capabilities and intrusion detection capabilities to all drones introduced to public authorities. However, this would be unrealistic as it causes substantial cost increases. It will also be difficult to equip many commercial drones with advanced security features. Therefore we proposed the minimum security requirements in Table~\ref{tab:SecurityRequirements} that the commercial drones introduced to public authorities should fulfill to prevent the vulnerabilities listed in Table~\ref{tab:VulnerabilityCategory}.

To prevent the vulnerability type V-1, it is necessary to identify authorized users when authorizing control of the drone and when using critical functions (\#1). Additionally, the repeated access attempts to the drone that keep fails should be restricted to avoid abnormal authentication trials (\#2).

The critical information such as the account information and the encryption key should be protected from exposure in the app to prevent the vulnerability type V-2$\cdotp$3 (\#3). Also, sensitive information transferred through the network between the controller and the drone should be encrypted for confidentiality (\#4) using a secure algorithm (\#5).

By prohibit unintended access to external servers, the vulnerability type of V-5 should be protected. If necessary, access to the external network should be limited using the white list access control mechanism.

The vulnerability types of V-1$\cdotp$2$\cdotp$4 are related to the malicious behaviors that can be attempt while the drone is in operation. The simultaneous accesses to the drone by multiple controllers should be prohibited to prevent such vulnerability types.

To prevent the occurrence of V-4 type of vulnerability, it is essential to restrict to provide services and to open ports which do not be explicitly recognized by users in addition to proscribing executing the vulnerable application services (\#8). It should also be limited to the execution of arbitrary commands and commands beyond permission even if the operator is authorized to access the drone (\#9).

\section{Conclusion}

In the future, the application of drones will be expanded further and will be combined with various technologies to create synergy effects. In line with this trend, public authorities will also expand the adoption of drones and prefer relatively economical commercial products rather than through exclusively developed expensive products.

When drones introduced to public authorities lacks security, the damage is evident to national property and information. However, if the drone has many security features, such as electronic warfare capabilities and intrusion persistence capabilities, it will be able to respond to various threats but will increase costs.

In this paper, we proposed the minimum security requirements that commercial drones introduced and operated in public authorities regardless of their operating field and purpose by analyzing control apps of various commercial drones.

We expect that this study promotes not only increasing security levels of public authorities but also strengthening those of overall commercial drones.

\bibliographystyle{splncs04}

\begin{thebibliography}{1}

\bibitem{DKim2015}
D. Kim, S. Baik, and S. Baek, "Proposing required video processing functionalities of future-oriented army small size unit Unmaned Aerial Vehicle (UAV) and verifying the employment concepts," Korean Journal of Military Arts and Science, Vol. 71, No. 1, pp.109-128, February 2015.

\bibitem{JPark2018}
J. Park, "Market status and prospective of global commercial drones," Aviation issue No.15, Korea Aerospace Research Institute, 2018.

\bibitem{IoTFS2015}
Internet of Things Forum, "Security requirements for drone-based IoT services," IoTFS-0079, December 2015.

\bibitem{TTAK2017}
Telecommunication Technology Association, "Security requirements for drone-based services," TTAK.KO-12.0317, December 2017.

\bibitem{IJun2017}
I. Jun and D. Hong, "Vulnerability analysis of drones using Wi-Fi," Proceedings of the Korean Society of Computer Information Conference, Vol. 25, No. 1, pp.219-222, January 2017.

\bibitem{DHong2017}
D. Hong, "Drone hacking and malware," The Journal of The Korean Institute of Communication Sciences, Vol. 34, No. 10, pp.3-9, July 2017.

\bibitem{JValente2017}
J. Valente and A.A. Cardenas, "Understanding security threats in consumer drones through the lens of the discovery quadcopter family," In Proceedings of the 2017 Workshop on Internet of Things Security and Privacy, pp.31-36, November 2017.

\bibitem{KHartmann2013}
K. Hartmann and C. Steup, "The vulnerability of UAVs to cyber attacks - an approach to the risk assessment," 5th International Conference on Cyber Conflict, pp.1-23, June 2013.

\bibitem{WJeon2016}
W. Jeon, "An analysis of UAVs' security vulnerabilities and development of protection profile based on Common Criteria," Journal of Information Technology and Architecture, Vol. 13, No. 4, pp.663-672, December 2016.

\bibitem{DYoo2007}
 D. Y. Yoo, J. W. Shin, and J. Y. Choi, "Home-network security model in ubiquitous environment," In Proceedings of World Academy of Science, Engineering and Technology, Vol. 1, No. 12, December 2007.

\end{thebibliography}

\end{document}